\documentclass[journal,onecolumn]{IEEEtran}

\usepackage{amssymb}
\usepackage{caption2}
\usepackage{amsmath}
\usepackage{multirow}
\usepackage{amsthm}

\newcommand{\ma}{\mathcal}

\newcommand{\fr}{\frac}
\newcommand{\lc}{\lceil}
\newcommand{\rc}{\rceil}
\newcommand{\lf}{\lfloor}
\newcommand{\rf}{\rfloor}

\begin{document}

\title{A New Piggybacking Design for Systematic MDS Storage Codes}
\author{Chong Shangguan and Gennian Ge
\thanks{The research of G. Ge was supported by the National Natural Science Foundation of China under Grant Nos. 11431003 and 61571310.
}
\thanks{C. Shangguan is with the School of Mathematics, Zhejiang University,
Hangzhou 310027, China (e-mail: 11235061@zju.edu.cn).}
\thanks{G. Ge is with the School of Mathematical Sciences, Capital Normal University,
Beijing 100048, China (e-mail: gnge@zju.edu.cn). He is also with Beijing Center for Mathematics and Information Interdisciplinary Sciences, Beijing, 100048, China.}
\thanks{Copyright (c) 2014 IEEE. Personal use of this material is permitted. However, permission to use this material for any other purposes must be obtained from the IEEE by sending a request to pubs-permissions@ieee.org.}
}
\maketitle

\begin{abstract}
    Distributed storage codes 
    have important applications in the design of modern storage systems. In a distributed storage system, every storage node has a probability to fail and once an individual storage node fails, it must be reconstructed using data stored in the surviving nodes. Computation load and network bandwidth are two important issues we need to concern when repairing a failed node. The traditional maximal distance separable (MDS) storage codes have low repair complexity but high repair bandwidth. On the contrary, minimal storage regenerating (MSR) codes have low repair bandwidth but high repair complexity. Fortunately, the newly introduced piggyback codes combine the advantages of both ones.

    In this paper, by introducing a novel piggybacking design framework for systematic MDS codes, we construct a storage code whose average repair bandwidth rate, i.e., the ratio of average repair bandwidth and the amount of the original data, can be as low as $\fr{\sqrt{2r-1}}{r}$, which significantly improves the ratio $\fr{r-1}{2r-1}$ of the previous result. In the meanwhile, every failed systematic node of the new code can be reconstructed quickly using the decoding algorithm of an MDS code, only with some additional additions over the underlying finite field.
    This is very fast compared with the complex matrix multiplications needed in the repair of a failed node of an MSR code.

\end{abstract}


\section{Introduction}

Due to their reliability and efficiency in data storage, distributed storage systems have attracted a lot of attentions in the last decades. In a distributed storage system, the whole data is stored in a collection of storage nodes. These nodes are physically independent and connected by a network. Since every single node has a probability to fail, redundancy is introduced to ensure the reliability of the system. In the literature, there are two strategies to guarantee redundancy: replication and erasure coding. Intuitively, replication is simple but inefficient. On the contrary, erasure coding provides much better storage efficiency. Therefore, in order to handle massive amount of information, erasure coding techniques have been employed by many modern distributed storage systems, for example, Google Colossus \cite{google}, HDFS Raid \cite{raid}, Total Recall \cite{total}, Microsoft Azure \cite{azure} and OceanStore \cite{oceanstore}.

Once an individual storage node fails, it must be reconstructed using data stored in the surviving nodes. There are four parameters we need to concern when repairing a failed node, such as computation load, network bandwidth, disk I/O and the number of accessed disks. 
In the literature, most of the existing storage codes are optimal for only one of these four parameters, for example, MDS codes for computation load, regenerating codes for network bandwidth \cite{Dimakis2010Network} and local repairable codes for the number of accessed disks \cite{Gopalan2012On,TamoLRC}.
The primary concern of this paper is to optimize the first two parameters. We define the average repair bandwidth rate, $\gamma$, to be the ratio of average repair bandwidth and the amount of the original data. In what follows, we will briefly review the repair complexity and repair bandwidth of three classes of storage codes, namely, MDS storage codes, MSR codes and piggyback codes.

The MDS code is a widely-used class of erasure codes for data storage, see for example, \cite{Tamo2015Zigzag,Wang2012Long}. It is optimal in terms of the redundancy-reliability tradeoff. A $(k+r,k)$ MDS storage code consists of $k+r$ storage nodes, with the property that the original message can be recovered from any $k$ of the $k+r$ nodes. It can tolerate the failure of any $r$ nodes. This property is termed the MDS property. A node is called systematic if it stores parts of the original message without coding. A systematic MDS code is an MDS code such that the original message is stored in $k$ nodes in the uncoded form. The remaining $r$ nodes, which are called parity nodes, store the parity data of the $k$ systematic nodes. From a practical standpoint, it is preferred to have the systematic feature, since in normal cases, data can be read directly from the systematic nodes without performing decoding. Many practical considerations also require the storage codes to be high-rate, i.e., $r\ll k$. Therefore, the repair efficiency of a failed systematic node is of great importance in the design of a distributed storage system.

In the case of MDS storage codes, the repair of a failed storage node only involves addition and multiplication in finite fields, which leads to a reasonable computation load in the repair process. However, to repair a single failed node, an MDS storage code needs to download the whole of its original data. In other words, the average repair bandwidth rate of MDS storage codes, $\gamma_{MDS}$, equals 1.

In 2010, Dimakis et al. \cite{Dimakis2010Network} introduced the notion of regenerating codes to reduce the repair bandwidth of distributed storage systems, where the failed systematic node is repaired by downloading a same amount of data from each of the surviving nodes.The MSR code is one of the two most important regenerating codes. It maintains the MDS property and has an average repair bandwidth rate $\gamma_{MSR}=\fr{k+r-1}{rk}$, 
which gives $\gamma_{MSR}\approx\fr{1}{r}$ when $r\ll k$. $\gamma_{MSR}$ becomes much smaller than $\gamma_{MDS}$ as $r$ grows larger. 
However, a drawback of the MSR code is that its repair algorithm of a failed systematic node involves multiplications of matrices, whose computational complexity may be too high for existing storage systems.

It is of great interest to construct storage codes with the following features: the MDS property, low repair complexity and low repair bandwidth. Motivated by these expectations, the seminal papers \cite{Rashmi2013B,Rashmi2013A} presented a piggybacking framework to combine the advantages of MDS codes and MSR codes. The idea of piggybacking is to take multiple instances of some existing code and adds carefully designed functions of data from one instance onto the other. As a result, the piggyback codes described in \cite{Rashmi2013B} (see, Section 4 of \cite{Rashmi2013B}) not only preserve the low computational complexity of MDS codes but also have an average repair bandwidth rate $\gamma_{RSR}=\fr{r-1}{2r-3}\approx\fr{1}{2}<\gamma_{MDS}$. Since then, this new idea has been applied successfully by several researchers. In 2013, it was adopted in the design of new storage systems for Facebook \cite{facebook}. In 2015, Yang et al. \cite{yang2015systematic} employed the piggybacking strategy to design new MSR codes with almost optimal repair bandwidth for parity nodes. Kumar et al. \cite{kumar2015family} also used this technique to construct codes with low repair bandwidth and low repair complexity, at the cost of lower fault tolerance.

It is not hard to see that the performance of piggyback codes lies between that of MDS codes and that of MSR codes. The main purpose of this paper is to design a new piggybacking framework to further reduce the repair bandwidth of the systematic nodes of a storage code. Our design can produce a new systematic MDS storage code with average repair bandwidth rate as low as $\gamma_{NEW}=\fr{\sqrt{2r-1}}{r}$. Obviously, our result significantly improves $\gamma_{RSR}$ for almost all choices of $r$. Furthermore, compared with the relatively high computational complexity of MSR codes, the repair of a failed storage node of the new code only involves addition and multiplication in some finite field.

The rest of this paper is organised as follows. In Section \ref{sec2}, we briefly review the piggybacking framework introduced in \cite{Rashmi2013B}. Our new piggybacking design is presented in Section \ref{sec3}. In Section \ref{sec4}, we compare our new storage code with some existing ones. We pose two open problems in Section \ref{sec5} for further research.

\section{The piggybacking framework}{\label{sec2}}

We will introduce some terminology defined in \cite{Rashmi2013B}. Denote by $\mathbb{F}:=\mathbb{F}_q$ the underlying finite field, where $q$ is a power of some prime number. The piggybacking framework operates on an arbitrary existing code, which is termed the base code. Without loss of generality, we can assume that the base code is associated with $n$ encoding functions $\{f_i\}_{i=1}^n$ and stored in $n$ storage nodes. Consider $m$ instances of the base code, then the initial encoded system is as follows:

\begin{table}[h]
\centering
\begin{tabular}{|c|c|c|c|c|}
  \hline
  Node 1 & $f_1(a_1)$ & $f_1(a_2)$ & $\cdots$ & $f_1(a_m)$ \\\hline
   $\vdots$ & $\vdots$ & $\vdots$ & $\ddots$ & $\cdots$ \\\hline
  Node $n$ & $f_n(a_1)$ & $f_n(a_2)$ & $\cdots$ & $f_n(a_m)$ \\
  \hline
\end{tabular}
\end{table}

\noindent where $a_1,\ldots,a_m$ denote the messages encoded under the $m$ instances. For every $1\le i\le n$ and $2\le j\le m$, one can add an arbitrary value $g_{i,j}(a_1,\ldots,a_{j-1})$ to $f_i(a_j)$. 
Here the functions $g_{i,j}:\mathbb{F}^k\longrightarrow\mathbb{F},~1\le i\le n,~2\le j\le m$ are termed piggyback functions, which can be chosen arbitrarily. The values to be added are termed piggybacks. Therefore, the symbol stored in the $i$-th node (row) and $j$-th instance (column) is $f_i(a_j)+g_{i,j}(a_1,\ldots,a_{j-1})$. The resulting piggyback code is depicted in Table \ref{piggybacked1}. The first instance contains no piggybacks since such arrangement allows $a_1$ to be recovered directly using the decoding algorithm of the base code.

\begin{table}[h]
\caption{The piggyback code}\label{piggybacked1}
\centering
\begin{tabular}{|c|c|c|c|c|}
  \hline
  Node 1 & $f_1(a_1)$ & $f_1(a_2)+g_{1,2}(a_1)$ & $\cdots$ & $f_1(a_m)+g_{1,m}(a_1,\ldots,a_{m-1})$ \\\hline
   $\vdots$ & $\vdots$ & $\vdots$ & $\ddots$ & $\cdots$ \\\hline
  Node $n$ & $f_n(a_1)$ & $f_n(a_2)+g_{n,2}(a_1)$ & $\cdots$ & $f_n(a_m)+g_{n,m}(a_1,\ldots,a_{m-1})$ \\
  \hline
\end{tabular}
\end{table}

In this paper, we take the base code to be a systematic $(k+r,k)$ MDS code, whose structure is described in Table \ref{MDS1},

\begin{table}[h]
\caption{The systematic $(k+r,k)$ MDS code}\label{MDS1}
\centering
\begin{tabular}{|c|c|c|c|c|}
  \hline
  Node 1 & $a_{1,1}$ & $a_{1,2}$ & $\cdots$ & $a_{1,m}$ \\\hline
  $\vdots$ & $\vdots$ & $\vdots$ & $\ddots$ & $\ddots$ \\\hline
  Node $k$ & $a_{k,1}$ & $a_{k,2}$ & $\cdots$ & $a_{k,m}$ \\\hline
  Node $k+1$ & $f_1(a_1)$ & $f_1(a_2)$ & $\cdots$ & $f_1(a_m)$ \\\hline
  $\vdots$ & $\vdots$ & $\vdots$ & $\ddots$ & $\vdots$ \\\hline
  Node $k+r$ & $f_r(a_1)$ & $f_r(a_2)$ & $\cdots$ & $f_r(a_m)$ \\
  \hline
\end{tabular}
\end{table}

 \noindent where we also take $m$ instances of the base code and denote $a_i=(a_{1,i},a_{2,i},\ldots,a_{k,i})^T$ for every $1\le i\le m$. The functions $\{f_i:1\le i\le r\}$ are called parity functions, which are chosen to ensure the MDS property of the code. The original data $\{a_1,a_2,\ldots,a_m\}$ is stored in the $k$ systematic nodes in the uncoded form. We can assume that every symbol in the array stores a unit amount of data. According to the piggybacking framework introduced in Table \ref{piggybacked1}, the systematic MDS code of Table \ref{MDS1} has the piggybacked form described in Table \ref{piggbackedmds}.

\begin{table}[h]
\caption{The piggybacked systematic $(k+r,k)$ MDS code}\label{piggbackedmds}
\centering
\begin{tabular}{|c|c|c|c|c|}
  \hline
  Node 1 & $a_{1,1}$ & $a_{1,2}$ & $\cdots$ & $a_{1,m}$ \\\hline
  $\vdots$ & $\vdots$ & $\vdots$ & $\ddots$ & $\ddots$ \\\hline
  Node $k$ & $a_{k,1}$ & $a_{k,2}$ & $\cdots$ & $a_{k,m}$ \\\hline
  Node $k+1$ & $f_1(a_1)$ & $f_1(a_2)+g_{1,2}(a_1)$ & $\cdots$ & $f_1(a_m)+g_{1,m}(a_1,\ldots,a_{m-1})$ \\\hline
  $\vdots$ & $\vdots$ & $\vdots$ & $\ddots$ & $\vdots$ \\\hline
  Node $k+r$ & $f_r(a_1)$ & $f_r(a_2)+g_{r,2}(a_1)$ & $\cdots$ & $f_r(a_m)+g_{r,m}(a_1,\ldots,a_{m-1})$ \\
  \hline
\end{tabular}
\end{table}

A crucial point in the addition of the piggybacks is that the functions $g_{i,j}$ can only operate on the message symbols of previous instances, namely $\{a_1,\ldots,a_{j-1}\}$. In the sequel we will call this constraint ``the piggybacking condition". It has been shown that such condition maintains the MDS property of an MDS code (see Theorem 1 and Corollary 2 of \cite{Rashmi2013B} for the details). In \cite{Rashmi2013B}, the authors presented several code constructions. The second one is the most efficient one in terms of repair bandwidth, whose minimal average repair bandwidth rate is $\gamma_{RSR}\ge\fr{r-1}{2r-3}$ and the equality holds when $r-1\mid k$. In their construction, they took $m:=2r-3$ instances of the base code. The function $f_i$ was defined to be $f_i(x)=<p_i,x>$ for $1\le i\le r$, where $p_i\in\mathbb{F}^k$ and $<\cdot,\cdot>$ denotes the conventional inner product over $\mathbb{F}$. 
Table \ref{piggbackedold} briefly describes the symbols stored in the parity node $k+i$, $i\in\{2,\ldots,r\}$. The variables $v_i$, $q_{i,j}$, $i\in\{2,\ldots,r-2\}$, $j\in\{1,\ldots,r-1\}$, involved in the computation of the piggybacks all belong to $\mathbb{F}^k$. Explicit expressions are not given here for the sake of saving space. At the first sight, it is likely to find this construction 
a bit complex and not easy to understand. In the next section, we will present a new piggybacking design which looks much cleaner 
and has an average repair rate as low as $\gamma_{NEW}=\fr{\sqrt{2r-1}}{r}$. 

Another family of piggyback codes was introduced in \cite{kumar2015family}. It was based on two classes of parity symbol such that the first class is used for good fault tolerance and the second class is used for reducing repair bandwidth and complexity. However, such construction does not maintain the MDS property. In Section \ref{sec4}, we will compare our construction with these codes.

\begin{table}[h]
\caption{The RSR piggyback code}\label{piggbackedold}
\centering
\begin{tabular}{|c|c|c|c|c|c|c|c|c|c|c|}
  \hline
 $p_i^Ta_1$ & $\cdots$ & $p_i^Ta_{r-2}$ & $q_{i,i-1}^Ta_{r-1}-\sum_{j=r}^{2r-3}p_i^Ta_j$ & $p_i^Ta_r+q_{i,1}^Tv_i$ & $\cdots$ & $p_i^Ta_{r+i-3}+q_{i,i-2}^Tv_i$ & $p_i^Ta_r+q_{i,i}^Tv_i$ & $\cdots$ & $p_i^Ta_{2r-3}+q_{i,r-1}^Tv_i$ \\
  \hline
\end{tabular}
\end{table}

\section{The new piggybacking design}{\label{sec3}}

In this section, we will present our new piggybacking design and the corresponding repair algorithm. Our main contribution is on the reduction of the repair bandwidth for systematic nodes, which is the primary concern of many existing storage codes.
Our design is based on an elaborative selection and placement of the piggybacking functions. We first begin with an example to illustrate our idea.


\subsection{The piggybacked (11,6) MDS code}

In this subsection, we describe in detail the piggybacking design and repair algorithm for an (11,6) systematic MDS code. Keep in mind the structure of a systematic MDS code described in Table \ref{piggbackedmds}. 
We will take 5 instances (we typically choose the number of instances equal to $r$) of the base code. The construction is described in Table \ref{example1}.

\begin{table}[h]
\caption{The piggybacked (11,6) MDS code}\label{example1}
\centering
\begin{tabular}{|c|c|c|c|c|}
  \hline
  $a_{1,1}$ & $a_{1,2}$ & $a_{1,3}$ & $a_{1,4}$ & $a_{1,5}$ \\\hline
  $a_{2,1}$ & $a_{2,2}$ & $a_{2,3}$ & $a_{2,4}$ & $a_{2,5}$ \\\hline
  $a_{3,1}$ & $a_{3,2}$ & $a_{3,3}$ & $a_{3,4}$ & $a_{3,5}$ \\\hline
  $a_{4,1}$ & $a_{4,2}$ & $a_{4,3}$ & $a_{4,4}$ & $a_{4,5}$ \\\hline
  $a_{5,1}$ & $a_{5,2}$ & $a_{5,3}$ & $a_{5,4}$ & $a_{5,5}$ \\\hline
  $a_{6,1}$ & $a_{6,2}$ & $a_{6,3}$ & $a_{6,4}$ & $a_{6,5}$ \\\hline
  $f_1(a_1)$ & $f_1(a_2)$ & $f_1(a_3)$ & $f_1(a_4)$ & $f_1(a_5)$ \\\hline
  $f_2(a_1)$ & $f_2(a_2)$ & $f_2(a_3)+a_{5,1}+a_{6,1}$ & $f_2(a_4)+a_{3,1}+a_{4,1}$ & $f_2(a_5)+a_{1,1}+a_{2,1}$ \\\hline
  $f_3(a_1)$ & $f_3(a_2)$ & $f_3(a_3)+a_{5,2}+a_{6,2}$ & $f_3(a_4)+a_{3,2}+a_{4,2}$ & $f_3(a_5)+a_{1,2}+a_{2,2}$ \\\hline
  $f_4(a_1)$ & $f_4(a_2)$ & $f_4(a_3)$ & $f_4(a_4)+a_{3,3}+a_{4,3}$ & $f_4(a_5)+a_{1,3}+a_{2,3}$ \\\hline
  $f_5(a_1)$ & $f_5(a_2)$ & $f_5(a_3)$ & $f_5(a_4)$ & $f_5(a_5)+a_{1,4}+a_{2,4}$ \\
  \hline
\end{tabular}
\end{table}

One can observe that all systematic nodes are partitioned into three subsets $S_1=\{1,2\}$, $S_2=\{3,4\}$ and $S_3=\{5,6\}$. Parts of symbols of $S_1,~S_2,~S_3$ are piggybacked in instances 5, 4 and 3, respectively. To be more precise, the symbols of the first four instances of $S_1$ are piggybacked in instance 5, the symbols of the first three instances of $S_2$ are piggybacked in instance 4 and the symbols of the first two instances of $S_3$ are piggybacked in instance 3. Consequently, nodes in different $S_i$ have different repair algorithms, we take one node of each $S_i$ as examples:
\begin{itemize}
  \item [(a)] Consider the repair of node 1. First $\{a_{i,5}:2\le i\le 6\}$ and $f_1(a_5)$ are downloaded and the entire vector $a_5$ is decoded using the MDS property. Then $\{f_{j+1}(a_5)+a_{1,j}+a_{2,j}:1\le j\le 4\}$ and $\{a_{2,j}:1\le j\le 4\}$ are downloaded from instance (column) 5 and node 2, respectively. Since $a_5$ is completely known, one can compute $\{f_{j+1}(a_5):1\le j\le 4\}$. Thus for $1\le j\le 4$, $a_{1,j}$ can be recovered by subtracting $a_{2,j}$ and $f_{j+1}(a_5)$ from $f_{j+1}(a_5)+a_{1,j}+a_{2,j}$. The total downloaded data in the repair of node 1 is $6+4\times 2=14$. The repair strategy of node 2 is similar.

  \item [(b)] Consider the repair of node 3. First $a_{3,5}$ is recovered by downloading $\{a_{i,5}:1\le i\le 6,~i\neq 3\}$ and $f_1(a_5)$ (using the MDS property). Then $a_{3,4}$ is recovered by downloading $\{a_{i,4}:1\le i\le 6,~i\neq 3\}$ and $f_1(a_4)$. It remains to recover $\{a_{3,j}:1\le j\le 3\}$. We will use the piggybacks added in instance (column) 4.  $\{f_{j+1}(a_4)+a_{3,j}+a_{4,j}:1\le j\le 3\}$ and $\{a_{4,j}:1\le j\le 3\}$ are downloaded from instance 4 and node 4, respectively. Since $a_4$ is completely known, one can compute $\{f_{j+1}(a_4):1\le j\le 3\}$. Thus for $1\le j\le 3$, $a_{3,j}$ can be recovered by subtracting $a_{4,j}$ and $f_{j+1}(a_4)$ from $f_{j+1}(a_4)+a_{3,j}+a_{4,j}$. The total downloaded data in the repair of node 3 is $6\times2+3\times 2=18$. The repair strategy of node 4 is similar.

  \item [(c)] Consider the repair of node 5. First $\{a_{5,j}:3\le j\le 5\}$ is recovered by downloading $\{a_{i,j}:1\le i\le 6,~i\neq 5,~3\le j\le 5\}$ and $\{f_1(a_j):3\le j\le 5\}$ (using the MDS property). It remains to recover $a_{5,1}$ and $a_{5,2}$, which can be done using the piggybacks added to $f_2(a_3)$ and $f_3(a_3)$. One can compute that the total downloaded data in the repair of node 5 is $6\times3+2\times 2=22$. The repair strategy of node 6 is similar.
\end{itemize}

It is easy to see that the proposed code has an average repair bandwidth $\fr{14+18+22}{3}=18$ and an average repair bandwidth rate $\gamma=\fr{18}{30}=\fr{3}{5}$. The amount of data required for the repair of nodes from different subsets lies in different hierarchies. The reason is that due to the piggybacking condition introduced in Section \ref{sec2}, the symbols stored in the latter instances can not be added as piggybacks onto the parity symbols of the former instances. Therefore, more information, which can only be obtained by the MDS property rather than piggybacking, will be needed when recovering symbols stored in the latter instances. For example, during the repair of node 1, we use the MDS property for only one time (to recover $a_5$), but in order to recover node 3 we have to use the MDS property twice (one time to recover $a_4$ and another time to recover $a_5$, since the information of $a_5$ can only be got from instance 5 using the MDS property). This observation indeed reveals the key idea of our construction: divide the systematic nodes into several subsets and piggyback symbols in the same subset onto same instance.


\subsection{The general piggybacking framework}

We will introduce our general piggybacking framework for the repair of the systematic nodes of an MDS code.
Take an arbitrary systematic $(k+r,k)$ MDS code as the base code. Generally speaking, to form the piggyback code $\ma{C}$, we will take $r$ instances of the base code. Let $\ma{S}=\{s_i:1\le i\le t\}$ be a set of $t$ positive integers such that $\sum_{i=1}^t s_i=k$. As shown in the above example, the $k$ systematic nodes of $\ma{C}$ are partitioned into $t$ groups, $\ma{S}_1,\ldots,\ma{S}_t$, such that $|\ma{S}_i|=s_i$ for $1\le i\le t$. Without lose of generality, we can assume that $\ma{S}_1=\{1,2,\ldots,s_1\}$ and $\ma{S}_i=\{\sum_{j=1}^{i-1}s_j+1,\ldots,\sum_{j=1}^{i}s_j\}$ for $2\le i \le t$. 
For a vector $\Lambda=(\lambda_1,\ldots,\lambda_k)$ of length $k$ over $\mathbb{F}$, the $t$ piggyback functions $\{g_i:1\le i\le t\}$ are defined to be $g_i(\Lambda)=\sum_{j\in\ma{S}_i}\lambda_j$. Our general piggybacking framework is presented as follows:

\begin{table}[h]
\caption{The general piggybacking framework}\label{general}
\centering
\begin{tabular}{|c|c|c|c|c|c|c|c|}
  \hline
  Node 1       & $a_{1,1}$      & $\cdots$ & $a_{1,r-t}$        & $a_{1,r-t+1}$             & $\cdots$ & $a_{1,r-1}$ & $a_{1,r}$\\\hline

  $\vdots$     & $\vdots$       & $\ddots$ & $\vdots$           & $\vdots$                  & $\ddots$ & $\vdots$ & $\vdots$\\\hline

  Node $k$     & $a_{k,1}$      & $\cdots$ & $a_{k,r-t}$        & $a_{k,r-t+1}$             & $\cdots$ & $a_{k,r-1}$ & $a_{k,r}$\\\hline

  Node $k+1$   & $f_1(a_1)$     & $\cdots$ & $f_1(a_{r-t})$     & $f_1(a_{r-t+1})$          & $\cdots$ & $f_1(a_{r-1})$ &$f_1(a_r)$\\\hline

  Node $k+2$   & $\vdots$     & $\ddots$ & $\vdots$     & $f_2(a_{r-t+1})+g_t(a_1)$          & $\cdots$ & $f_2(a_{r-1})+g_2(a_1)$ &$f_2(a_r)+g_1(a_1)$\\\hline

  $\vdots$     & $\vdots$       & $\ddots$ & $\vdots$           & $\vdots$ & $\ddots$ & $\vdots$ & $\vdots$\\\hline

  Node $k+r-t+1$ & $\vdots$ & $\ddots$ & $\vdots$ & $f_{r-t+1}(a_{r-t+1})+g_t(a_{r-t})$                  & $\ddots$ & $\vdots$ & $\vdots$\\\hline

  $\vdots$     & $\vdots$       & $\ddots$ & $\vdots$           & $\vdots$ & $\ddots$ & $\vdots$ & $\vdots$\\\hline

  Node $k+r-1$ & $\vdots$ & $\ddots$ & $\vdots$ & $\vdots$                  & $\ddots$ & $f_{r-1}(a_{r-1})+g_2(a_{r-2})$ & $\vdots$\\\hline

  Node $k+r$   & $f_r(a_1)$     & $\cdots$ & $f_r(a_{r-t})$     & $f_r(a_{r-t+1})$          & $\cdots$ & $f_r(a_{r-1})$ & $f_r(a_r)+g_1(a_{r-1})$\\
  \hline
\end{tabular}
\end{table}

Note that in the above table, the first $k+1$ nodes remain unchanged.
Our construction can be summarized as follows:
\begin{itemize}
  \item [(a)] None of the symbols of $a_r$ are piggybacked.
  \item [(b)] For $r-t+1\le j\le r-1$, the symbols of $a_j$ belonging to $\cup_{l=1}^{r-j}\ma{S}_l$ are piggybacked. More precisely, for $1\le l\le r-j$, the symbols of $a_j$ restricted to $\ma{S}_l$ are piggybacked in the $(j+1)$-th parity node of instance $r-l+1$. 
  \item [(c)] For $1\le j\le r-t$, all symbols of $a_j$ are piggybacked. More precisely, for $1\le l\le t$, the symbols of $a_j$ restricted to $\ma{S}_l$ are piggybacked in the $(j+1)$-th parity node of instance $r-l+1$. 
\end{itemize}

Consequently, the amount of data required for the repair of nodes from different subsets also lies in different hierarchies. For example, assume that we want to recover some failed node $i$ of $\ma{S}_l$, say, the symbols $\{a_{i,j}:1\le j\le r\}$. Note that for $1\le j\le r-l$, the symbol $a_{i,j}$ is piggybacked in the form $f_{j+1}(a_{r-l+1})+g_l(a_j)$ in the $(j+1)$-th parity node of instance $r-l+1$. One can recall the $i$-th row and the $(r-l+1)$-th column of Table \ref{general} for a better understanding:

\begin{center}
\begin{tabular}{|c|c|c|c|c|c|c|}
  \hline
           & &          &    $\vdots$                   &   &          &  \\\hline
  $a_{i,1}$ &$\cdots$& $a_{i,r-l}$  & $a_{i,r-l+1}$      & $a_{i,r-l+2}$     & $\cdots$ & $a_{i,r}$  \\\hline
            &&          & $f_1(a_{r-l+1})$         &    &      &   \\\hline
            &&          & $f_2(a_{r-l+1})+g_l(a_1)$ & & &  \\\hline
   &  & &$\vdots$ &  &  &\\\hline
   &  & &$f_{r-l+1}(a_{r-l+1})+g_l(a_{r-l})$ &  &  & \\\hline
   &  & &$f_{r-l+2}(a_{r-l+1})$ &  &  &\\\hline
   &&&$\vdots$&&&\\\hline
   &&&$f_r(a_{r-l+1})$&&&\\
  \hline
\end{tabular}
\end{center}

\noindent To recover $\{a_{i,j}:1\le j\le r\}$, firstly, each $a_{i,j}$, $r-l+1\le j\le r$ can only be reconstructed using the MDS property, hence the amount of data needed to be downloaded is $kl$. Secondly, each $a_{i,j}$, $1\le j\le r-l$ can be reconstructed by downloading $f_{j+1}(a_{r-l+1})+g_l(a_j)$ and $\{a_{i',j}:i'\in\ma{S}_l,~i'\neq i\}$, hence the amount of data needed to be downloaded is $(r-l)|\ma{S}_l|$. Since $f_{j+1}(a_{r-l+1})$ is known after $a_{r-l+1}$ is recovered, $a_{i,j}$ can be reconstructed by subtracting $f_{j+1}(a_{r-l+1})$ and $\sum_{i':i'\in\ma{S}_l,i'\neq i} a_{i',j}$ from $f_{j+1}(a_{r-l+1})+g_l(a_j)$. Therefore, the amount of data needed to be downloaded in the repair of node $i\in\ma{S}_l$ is $kl+(r-l)s_l$. We can conclude that the total amount of data needed to be downloaded in the repair of all systematic nodes is

\begin{equation*}
    \begin{aligned}
       \sum_{l=1}^t (kl+(r-l)s_l)s_l.
    \end{aligned}
\end{equation*}

\noindent Now it remains to find the minimal value of (\ref{piggy1}):

\begin{equation}\label{piggy1}
    \begin{aligned}
       \min&\sum_{l=1}^t (kl+(r-l)s_l)s_l,\\
       s.t.& ~\sum_{l=1}^t s_l=k~and~s_1,\ldots,s_t,t\in\mathbb{Z}^+.
    \end{aligned}
\end{equation}

Unfortunately, we are not able to compute the minimum value of (\ref{piggy1}) exactly. 
One may apply the Lagrange multiplier method to get a minimal value of (\ref{piggy1}) for every appropriate choice of $t$ and end up with a function of $t$, then compute the minimum of this function. We have tried along this line but found the computation to be too involved. Nevertheless, we can always let $s_i$ be some special values such that the target function is small enough. For instance, we can set $s_1=\cdots=s_t=\fr{k}{t}$, which leads to an average repair bandwidth rate

\begin{equation}\label{piggy2}
    \begin{aligned}
       \gamma&=\fr{1}{rk^2}\sum_{l=1}^t \fr{k}{t}(kl+(r-l)\fr{k}{t})\\
       &=\fr{1}{2}(\fr{t}{r}+\fr{1}{t}(2-\fr{1}{r})).\\
    \end{aligned}
\end{equation}

\noindent By the mean value inequality, (\ref{piggy2}) attains its minimum $\gamma=\fr{\sqrt{2r-1}}{r}$ when $t=\sqrt{2r-1}$.

We denote the code with above parameters by $\ma{C}_{NEW}$. If we repair every failed parity node simply by downloading the whole original data, then provided $r\ll k$, the average repair bandwidth rate for all storage nodes will be $$\gamma_{NEW}=\fr{\fr{\sqrt{2r-1}}{r}k+r}{k+r}\approx\fr{\sqrt{2r-1}}{r}=\ma{O}(\fr{1}{\sqrt{r}}).$$

In the remaining of this section we will present a slight improvement of the above construction. Note that in Table \ref{general}, for $r-t+1\le j\le r$, the number of symbols added to the parity nodes of instance $j$ is $(j-1)s_{r-j+1}$. One can see that these symbols are piggybacked in $j-1$ parity symbols of instance $j$ and there are still $r-j$ parity symbols leaving unused (the first parity symbol is used to ensure the MDS repair of $a_j$). Indeed, adding the $(j-1)s_{r-j+1}$ symbols as evenly as possible to all $r-1$ available parity nodes will lead to a better repair bandwidth. As an example, we reformulate the piggybacked pattern of Table \ref{example1} into the following one:

\begin{table}[h]
\centering
\begin{tabular}{|c|c|c|}
  \hline

   $f_2(a_3)+a_{5,1}$ & $f_2(a_4)+a_{3,1}$ & $f_2(a_5)+a_{1,1}+a_{2,1}$ \\\hline
   $f_3(a_3)+a_{5,2}$ & $f_3(a_4)+a_{3,2}+a_{4,2}$ & $f_3(a_5)+a_{1,2}+a_{2,2}$ \\\hline
   $f_4(a_3)+a_{6,1}$ & $f_4(a_4)+a_{3,3}+a_{4,3}$ & $f_4(a_5)+a_{1,3}+a_{2,3}$ \\\hline
   $f_5(a_3)+a_{6,2}$ & $f_5(a_4)+a_{4,1}$ & $f_5(a_5)+a_{1,4}+a_{2,4}$ \\
  \hline
\end{tabular}
\end{table}

Compared with the previously described construction, the recover of each symbol of $\{a_{5,1},a_{5,2},a_{6,1},a_{6,2},a_{3,1},a_{4,1}\}$ requires less information (2 instead of 3). In general, similar improvements can be made for every $r-t+1\le j\le r-1$. As in the piggybacking design of \cite{Rashmi2013B}, for instance $j$, one may divide all symbols piggybacked in that instance into $r-1$ groups. Let $t_f=\lf\fr{(j-1)s_{r-j+1}}{r-1}\rf$, $t_c=\lc\fr{(j-1)s_{r-j+1}}{r-1}\rc$, $t=(j-1)s_{r-j+1}-(r-1)t_f$. The first $t$ groups are chosen of size $t_c$ each and the remaining $r-1-t$ groups have size $t_f$ each. Then the systematic symbols involved in the sum $g_{r-j+1}(a_1)+\cdots+g_{r-j+1}(a_{j-1})$ are added ``evenly" to these $r-1$ parity nodes. This new placement of symbols will sightly reduce the repair bandwidth. We will not carry out an explicit computation here since such reduction only affects the constant coefficient before $\fr{1}{\sqrt{r}}$.


We would like to comment that in the practical setting, when the values $k$ and $r$ are given, one may do better than $\fr{\sqrt{2r-1}}{r}$ by either carrying out a more careful optimization of (\ref{piggy1}) or applying the trick on the placement of the piggybacks discussed above.

\section{Comparison of some existing codes}{\label{sec4}}

In this section, we compare the performance of some storage codes, namely, MDS code, the old piggyback codes introduced in \cite{kumar2015family,Rashmi2013B} and the one newly constructed in Section \ref{sec3}. In order to evaluate the repair complexity and encoding complexity of these codes, we first consider the complexity of elementary arithmetic operations of elements in the underlying finite field, which is denoted by $\mathbb{F}_q$. Denote $e=\lc\log_2 q\rc$, then an addition requires $e$ and a multiplication requires $e^2$ times of elementary binary additions, respectively. For $1\le i\le r$, we define the parity functions $f_i$ to be $f_i(a)=<p_i,a>$ for carefully chosen vectors $p_i\in\mathbb{F}_q^k$. Then the repair complexity of a single node of an MDS code (with only one instance) is $ke^2+(k-1)e$. We set $x:=ke^2+(k-1)e$ for convenience. Consider the new piggyback code, for $1\le l\le t$, the repair complexity of a systematic node $i\in\ma{S}_l$ is
$$lx+(r-l)(x+s_le)=rx+(r-l)s_le,$$
\noindent where the first part of the summation corresponds to the computation cost of repairing the last $l$ symbols in node $i$, and the second part of the summation corresponds to the computation cost of repairing the first $r-l$ symbols in node $i$.
The total computation cost of repairing all systematic nodes is $\sum_{l=1}^t s_l(rx+(r-l)s_le)$. On the other hand, the computation cost of repairing all $r$ parity nodes is $r^2x+\sum_{l=1}^t (r-l)s_le$. Thus the average repair complexity of all $k+r$ nodes of the new code is at most
$$\fr{\sum_{l=1}^t s_l(rx+(r-l)s_le)+r^2x+\sum_{l=1}^t (r-l)s_le}{k+r}=rx+\fr{\sum_{l=1}^t(r-l)s_l(s_l+1)e}{k+r}.$$

One can also compute the encoding complexity of an MDS code (with only one instance) and the new code, which is $rx$ and $r^2x+\sum_{l=1}^t s_l(r-l)e$, 
respectively. It is easy to check that the repair complexity and encoding complexity of the new code are very close to those of an MDS code with $r$ instances, which are $rx$ and $r^2x$, respectively. One can also determine the corresponding parameters of the piggyback codes constructed in \cite{kumar2015family,Rashmi2013B}. We can summarize the computation results as follows:

\begin{table}[h]
\caption{Comparison of some $(k+r,k)$ piggyback codes}
\centering
\begin{tabular}{|c|c|c|c|c|c|}
  \hline
  & Number of Instances &Fault Tolerance & Average Repair Bandwidth Rate & Average Repair Complexity & Encoding Complexity \\\hline
  MDS & 1 & $r$ & 1 & $x$ & $rx$ \\\hline
  RSR  \cite{Rashmi2013B} & $2r-3$ & $r$ & $\fr{r-1}{2r-3}$ & $\ma{O}((2r-3)x)$ & $\le(2r-3)rx+kre^2+kre$\\\hline
  KAAB \cite{kumar2015family} & $k$ & $\ge n_{A}-k-\tau+1$ & $<\fr{k+\tau+(k-\tau-1)^2}{k^2}$ & $\fr{C_R}{k}$ & $C_E$\\\hline
  New code & $r$ & $r$ & $\approx\fr{\sqrt{2r-1}}{r}$ & $\le rx+\sqrt{r}x$ & $\le r^2x+kre$ \\\hline
\end{tabular}
\end{table}

\noindent where $n_A\le\min\{k+r,2k\}$, $\tau\ge1$, $C_R,~C_E$ are defined in \cite{kumar2015family} and we take the new code to be $\ma{C}_{NEW}$.

Taking the number of instances into account, we can find that the fault tolerance, average repair complexity and encoding complexity of MSR, RSR and the new code are very close. For the average repair bandwidth rate, it holds that $\gamma_{NEW}\ll\gamma_{RSR}<\gamma_{MDS}$. For the KAAB code, by \cite{kumar2015family}, its average repair complexity and encoding complexity can even be lower than the MDS code. However, if we set $\gamma_{KAAB}=\Theta(\fr{1}{\sqrt{r}})$, then we have $\tau=\Theta(k(1-\fr{1}{r^{1/4}})$, which leads to a dramatic loss on the fault tolerance. Another nice feature of our construction is that, the number of instances is less than that of RSR code and much less than that of KAAB code provided $r\ll k$.

As discussed in \cite{Rashmi2013B}, practical data centers require the storage codes to be MDS, high-rate, have a small number of instances, and of course, have  low repair bandwidth and low repair complexity. It is not hard to find that the newly constructed code fits these requirements much better than the other ones.

\section{Conclusion}{\label{sec5}}
The main purpose of this paper is to optimize the repair bandwidth and repair complexity in the repair of a failed node in the distributed storage systems. Unfortunately, in our view it seems very hard to construct a code satisfying the following three properties:
\begin{itemize}
  \item [(a)] the MDS property,
  \item [(b)] the repair complexity is close to that of an MDS code,
  \item [(c)] the average repair bandwidth rate is close to that of an MSR code, which can be as low as $c/r$ for some constant $c$.
\end{itemize}

\noindent The new code introduced in this paper only satisfies the first two requirements, and has an average repair bandwidth rate $\gamma_{NEW}=\Theta(\fr{1}{\sqrt{r}})$. Therefore, we would like to pose two open problems for further research:

\textbf{Open Problem 1.} Establish an equality or an inequality such that the tradeoff between repair complexity and repair bandwidth can be written down mathematically.

\textbf{Open Problem 2.} Under conditions (a) and (b), determine the minimal average repair bandwidth rate for the repair of a failed node.

\bibliographystyle{IEEEtranS}
\bibliography{reference}
\end{document}